\documentclass[reprint,prl,aps,superscriptaddress]{revtex4-2}
\usepackage{times}
\usepackage{graphicx}
\usepackage{amsmath, braket, amsfonts}
\usepackage{amssymb}
\usepackage{sidecap}
\usepackage{bm, color, ulem}
\usepackage{xcolor}
\usepackage{ragged2e}
\usepackage{wrapfig,lipsum,booktabs}
\usepackage{lineno}

\usepackage{siunitx}

\newcommand{\be}{\begin{equation}}
\newcommand{\ee}{\end{equation}}

\DeclareUnicodeCharacter{03BD}{{$\nu$}}

\makeatletter
\def\maketitle{
\@author@finish
\title@column\titleblock@produce
\suppressfloats[t]}
\makeatother

\begin{document}
\title{Fluctuating magnetism and Pomeranchuk effect in multilayer graphene}
\author{Ludwig Holleis}  
\affiliation{Department of Physics, University of California at Santa Barbara, Santa Barbara CA 93106, USA}
\author{Tian Xie}
\author{Siyuan Xu}
\author{Haoxin Zhou}
\author{Caitlin L. Patterson}
\affiliation{Department of Physics, University of California at Santa Barbara, Santa Barbara CA 93106, USA}
\author{Archisman Panigrahi}
\affiliation{Department of Physics, Massachusetts Institute of Technology, Cambridge, MA 02139, USA}
\author{Takashi Taniguchi}
 \affiliation{International Center for Materials Nanoarchitectonics,
 National Institute for Materials Science,  1-1 Namiki, Tsukuba 305-0044, Japan}
 \author{Kenji Watanabe}
 \affiliation{Research Center for Functional Materials,
 National Institute for Materials Science, 1-1 Namiki, Tsukuba 305-0044, Japan}
\author{Leonid S. Levitov}
\affiliation{Department of Physics, Massachusetts Institute of Technology, Cambridge, MA 02139, USA}
\author{Chenhao Jin}
 \affiliation{Department of Physics, University of California at Santa Barbara, Santa Barbara CA 93106, USA}
\author{Erez Berg} 
\affiliation{Department of Condensed Matter Physics, Weizmann Institute of Science, Rehovot 76100, Israel}
\author{Andrea F. Young}
\email{andrea@physics.ucsb.edu}
 \affiliation{Department of Physics, University of California at Santa Barbara, Santa Barbara CA 93106, USA}
\date{\today}

\begin{abstract}
\end{abstract}

\maketitle

\textbf{
Magnetism typically arises from the effect of exchange interactions on highly localized fermionic wave functions in f- and d-atomic orbitals. 
In rhombohedral multilayer graphene (RMG), in contrast, magnetism---manifesting as spontaneous polarization into one or more spin and valley flavors\cite{Shi2020,Zhou2021a,Zhou2022,Barrera2022,Seiler2022,Han2023,Zhang2023}---originates from itinerant electrons near a Van Hove singularity. 
Here, we show experimentally that the electronic entropy in this system shows signatures typically associated with disordered local magnetic moments, unexpected for electrons in a fully itinerant metal.
Specifically, we find a contribution $\Delta S\approx 1 k_B/$~charge carrier that onsets at the Curie temperature and survives over one order of magnitude in temperature.  First order phase transitions show an isospin ``Pomeranchuk effect'' in which the 
fluctuating moment phase is entropically favored over the nearby symmetric Fermi liquid\cite{Saito2021,Rozen2021}.  
Our results imply that despite the itinerant nature of the electron wave functions, the spin- and valley polarization of individual electrons are decoupled, a phenomenon typically associated with localized moments, as happens, for example, in solid $^3$He\cite{Pomeranchuk1950}. Transport measurements, surprisingly, show a finite temperature resistance minimum within the fluctuating moment regime, which we attribute to the interplay of fluctuating magnetic moments and electron phonon scattering. 
Our results highlight the universality of soft isospin modes to two-dimensional flat band systems.}

\section{Introduction}
Magnetism within localized electron orbitals is typically described within a Hubbard model, which accounts for the on-site Coulomb repulsion but also allows for inter-site hopping.  
Hubbard models capture several key aspects of many magnetic materials, correctly predicting the scale of the ordering temperature, $T_C$, and the Curie-Weiss behavior of the magnetic susceptibility in the high temperature disordered phase.  In contrast, only a handful of materials show magnetism in the complete absence of localized orbitals near the Fermi level\cite{Santiago2017}.  This limit is captured by the Stoner model, which accounts for the competition between kinetic energy and Coulomb repulsion but fails to quantitatively describe the disorder state\cite{Moriya1985}.

Moir\'e flat band systems from graphene and other two dimensional materials are a recent addition to the family of ferromagnetic materials constructed from nonmagnetic constituents\cite{Cao2018,Sharpe2019,Chen2019,Serlin2020,Lu2024,Cai2023,Zeng2023}.  
However, in these systems, the periodic superlattice potential endows the single-particle wave functions with a localized structure on the scale of the Fermi wavelength, allowing a Hubbard-like description to apply\cite{Wu2018,Song2022}. 
Indeed, anomalously large entropy\cite{Saito2021,Rozen2021} and magnetic susceptibility\cite{Tang2020} at high temperatures have been attributed to fluctuating local moments. Theoretical approaches based on the localization of isospins on the moir\'e lattice have successfully explained the implied decoupling of charge and magnetic degrees\cite{Song2022,Shi2022,Hu2023,Hu2023a,Datta2023,Chou2023,Lau2023,Zhou2024,Ledwith2024}.

Recently, magnetism and superconductivity were discovered in rhombohedral multilayer graphene (RMG) \cite{Shi2020,Zhou2021,Zhou2021a,Zhou2022,Barrera2022,Seiler2022,Zhang2023,Holleis2023,Han2023,Seiler2023,Li2024,Han2024,Lu2024,Arp2023}. 
Rhombohedral graphite is defined as having the ``ABCABC...'' stacking order; rhombohedral N-layer graphene consists of an N-layer subset of this stacking.  We note that in the bilayer case (`R2G'), rhombohedral and Bernal stacking (defined for graphite by the  ``ABABAB...'' stacking order) are identical.  Crucially, however,  RMG systems of all layer numbers, including R2G, share a similar electronic structure characterized by two low energy bands featuring a gate tunable van Hove singularity in the low energy density of states at charge densities $|n_e|\approx 10^{12} \mathrm{cm^{-2}}$. 
RMGs provide an ideal platform to study the potential role of itinerancy in the isospin dynamics of flat band systems in a continuum limit where neither the graphene lattice nor any moir\'e lattices play any role and lattice-based theoretical models are not expected to apply.

\section{Static and fluctuating magnetism in RMG}

\begin{figure*}
    \centering
    \includegraphics[width = \textwidth]{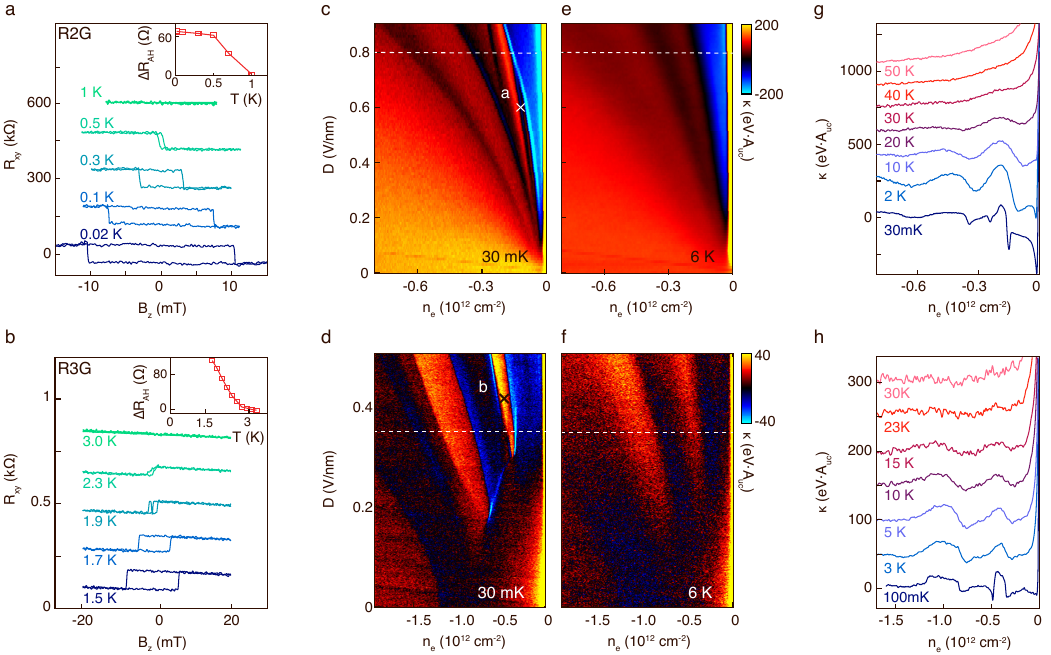}
    \caption{\textbf{Static and fluctuating magnetism in Bernal Bilayer Graphene (R2G) and Rhombohedral Trilayer  Graphene (R3G).} 
    \textbf{(a)}
    $R_{xy}$ in the quarter metal regime of R2G (see panel c) as a function of $B_z$ and for several different temperatures. $B_{||}$ = 250 mT for all data sets, which are offset for clarity. 
    The inset shows the magnitude of the anomalous Hall step, from which we extract the Curie temperature $T_C\approx$ 1 K for R2G. 
    \textbf{(b)} Analogous data for R3G, at the point indicated in panel d).  
    $T_{c,R3G}$ $\approx$ 3K, as shown in the inset.
    \textbf{(c)} Inverse electronic compressibility, $\kappa$,  measured at  30 mK for R2G and 
    \textbf{(d)} R3G.  
    \textbf{(e)} $\kappa$ measured at 6 K for R2G and  
    \textbf{(f)} R3G. 
    \textbf{(g)} $\kappa$ as a function of $n_e$ with fixed $D=$ 0.8 V/nm at different temperatures for R2G. 
    \textbf{(h)} $\kappa$ as a function of $n_e$ with fixed $D=$ 0.35 V/nm at different temperatures for R3G. Curves in panels g), h) are offset for clarity.  
    }
    \label{fig:1}
\end{figure*}

While superconductivity and magnetism are natural consequences of the high density of states in flat band systems, decoupled isospin moment behavior is not expected in a fully itinerant metal\cite{Santiago2017} where spatial overlap between single particle wave functions naively leads to exchange couplings that suppress moment fluctuations. As a result, the entropy due to fluctuating isospins above the Curie temperature is expected to be small. Here, we put these expectations to the test by measuring the temperature dependent thermodynamic and transport properties of rhombohedral 2- and 3-layer graphene (R2G and R3G, respectively).  
Fig. \ref{fig:1}a shows  $R_{xy}$ measured in R2G, in a regime of carrier density $n_e$ and applied perpendicular displacement field where the ground state is known to consist of a spin and valley polarized ferromagnet\cite{Zhou2022}. 
The anomalous Hall signal associated with long range valley order vanishes at $T_{C}$ $\approx$ 1K for R2G, which we define as the effective Curie temperature for the long range valley order. 
Analogous measurements for R3G (Fig. \ref{fig:1}b) show a slightly higher Curie temperature of $T_C\approx$ 3K.  
Low temperature measurements of the inverse compressibility, $\kappa$ = $\partial\mu$/$\partial n_e$, where $\mu$ is the chemical potential of the graphene multilayer, are shown in Figs. \ref{fig:1}c-d for R2G and R3G, respectively.  
At our base temperature of 30 mK, the $n_e$- and $D$-tuned phase diagrams show a series of phases with distinct values of $\kappa$, demarcated in many cases by sharp negative spikes (rendered in blue on the chosen color scale).  
These are associated with phases with one or more broken isospin symmetries, separated by first order transitions\cite{Zhou2021a, Barrera2022,Seiler2022}. 
Notably, increasing the temperature above $T_C$ (see $\kappa$ measurements at 6 K in Figs. \ref{fig:1}e-f) washes out the first order phase transitions but leaves the broad structure of the compressibility intact\cite{Barrera2022}; in particular, the compressibility shows a non-monotonic dependence on charge carrier density that is absent in the single particle electronic structure of either allotrope.  Modulations in $\kappa$ persist up to $T\approx 30 K$ for both systems, more than ten times the Curie temperature of long range valley order. We observe similar signatures using exciton sensing\cite{Xie2024} techniques, which are sensitive to the isospin polarization of the electrons near the Fermi level (see  Extended Data Fig. \ref{fig:S_optical} and Methods).

For both compressibility and exciton sensing, signals may arise in a fluctuating system in which the \textit{average} number of occupied isospin flavors changes with $n_e$ but the particular occupied flavor fluctuates on time scales much faster than those of the measurements, and on length scales much smaller than the probe area. We thus attribute the compressibility modulations to a fluctuating `local moment'  regime for $3K\lesssim T\lesssim30K$ in which the spin and valley flavor of nearly all of the electrons within the Fermi sea become uncorrelated.  These results are directly analogous to observations in twisted bilayer graphene (tBLG)\cite{Saito2021,Rozen2021,Xie2024}.    
%


\section{Isospin entropy and Pomeranchuk effect}

\begin{figure*}
    \centering
    \includegraphics[width = 7in]{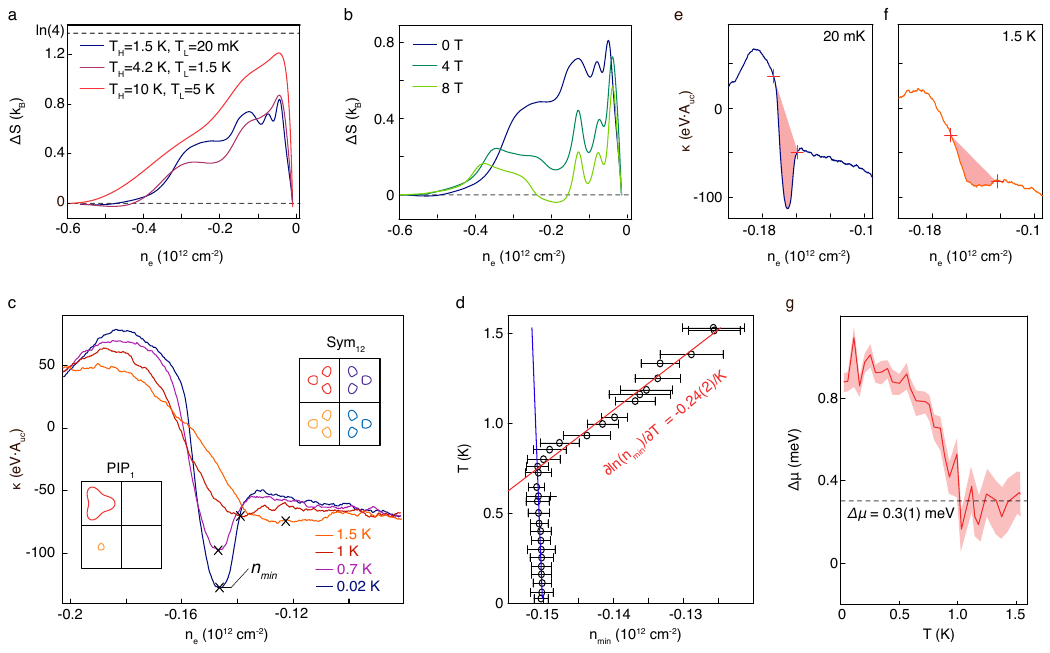}
    \caption{\textbf{Isospin entropy and Pomeranchuk effect in R2G.} 
\textbf{(a)} 
    Density dependent excess entropy per particle, $\Delta S$, for three temperature differences at $D$ = 0.5 V/nm. 
    The extraction of $\Delta S$ is described in detail in the methods and Fig. \ref{fig:S_DeltaS}. 
\textbf{(b)} 
    Suppression of $\Delta S$ with in-plane magnetic field. Here $\Delta S$ is measured with $T_H=1.5 K$ and $T_L=20 mK$. 
\textbf{(c)} $\kappa$ for $D$ = 0.8 V/nm and different temperatures in the vicinity of the transition from a low-$|n_e|$ symmetric state with twelve Fermi pockets to a spin- and valley polarized phase (see insets, and ref. \cite{Zhou2022}).  The `x' symbols mark the $\kappa$ minimum associated with the first order transition.  
\textbf{(d)} $n_{min}$, the position of the first order transition shown in panel c, as a function of $T$. The error bars are determined by the coefficient of determination from the quadratic fits used to determine the position of the minimum in $\kappa$. Above $T_C\approx 0.8K$, we find a logarithmic derivative 
$\partial \ln (n_{min})/\partial T$ = -0.24(2)/K.
\textbf{(e)} $\kappa$ at 20mK and 
\textbf{(f)} 1.5K.  To determine the chemical potential jump $\Delta \mu$ at the first order phase transition, we integrate $\kappa$  over the highlighted regions as described in details in methods and Fig. \ref{fig:S_Pomeranchuk}.  
\textbf{(g)} Temperature dependence of $\Delta \mu$. The shaded area serve as error bars determined as described in the methods section. The total entropy difference across the transition is given by $-\Delta \mu \cdot \partial \ln n_{min} /\partial T$ $\approx$ 0.8(3) k$_B$.}
\label{fig:3}
\end{figure*}


In tBLG, fluctuating moments manifest as a large contribution to the electronic entropy\cite{Saito2021,Rozen2021}, of order 1~k$_B$/carrier, which is suppressed by in-plane magnetic field if $g\mu_B B_\parallel\approx k_B T$. 
A striking consequence of this entropy is an isospin analogue of the Pomeranchuk effect\cite{Pomeranchuk1950}, in which raising the temperature favors a fluctuating moment phase over a fully spin and valley unpolarized Fermi liquid. 
To further explore the analogy between tBLG and crystalline RMG, Fig. \ref{fig:3}a shows the electronic entropy per electron, $\Delta S$, measured relative to charge neutrality for R2G at D=0.5V/nm. 
$\Delta S$ is measured via the Maxwell relation between the entropy and chemical potential, $\mu$ (see Methods). We observe a large excess entropy of approximately 0.8 k$_B$ per particle starting from temperatures as small as 1.5K (see also Fig. \ref{fig:S_smallerTsteps}). The entropy is large for $-0.35 \times 10^{12} \mathrm{cm^{-2}}<n_e<-0.05\times 10^{12} \mathrm{cm^{-2}}$ and is approximately displacement field independent (Fig. \ref{fig:S_smallerTsteps}c), corresponding to the range where the low-temperature states show broken isospin symmetry.

In the absence of isospin order, the electrons are expected to form a Fermi liquid whose degeneracy suppresses the entropy per particle, with $\Delta S\approx \frac{1}{n_e}\frac{\pi^2}{6}\rho k_B^2 T$ in the low temperature limit (here $\rho$ is the electronic density of states); this entropy is expected to be negligible in RMG at 2K (see Fig. \ref{fig:S_DeltaS}).
The low temperature entropy of a state with long range isospin order---such as a quarter metal with a single Fermi sea---may contain an additional contribution due to  thermal population of low-energy collective modes but is also expected to be small.  
We ascribe the observed excess entropy $\Delta S\sim k_B$  to the decoupling of isospins of the individual conduction electrons; with each electron having arbitrary and independent polarization in the isospin space but with the volume of momentum space occupied similar between the ordered and disordered state.
A direct consequence of this picture is a large magnetic susceptibility, $\chi$, above $T_C$.  We measure  $\chi \approx$ 0.15 $\mu_B/T$  (see Fig. \ref{fig:S_DeltaM}), which exceeds the Pauli susceptibility expected from the single-particle picture by at least an order of magnitude.

Further evidence of the isospin origin of the entropy and magnetic susceptibility above $T_C$ is shown in Fig. \ref{fig:3}b, with $\Delta S$ measured for several different applied in-plane magnetic fields.  
Entropy is suppressed by the in-plane magnetic field, with a near complete suppression for $n_e\approx -0.2 \cdot 10^{12}\mathrm{cm^{-2}}$.  
This suppression is correlated with the isospin structure of the symmetry-broken phases at low temperatures at the same densities.  
In RMG, isospin order may correspond to spin polarization, valley polarization, or both.  
Allowing for fluctuations in this order at high temperature, a spin and valley polarized state with a single Fermi surface could show excess entropy as high as $\ln(4)$k$_B$ corresponding to a random orientation of both spin and valley for each carrier.  
In this case, an in-plane magnetic field would suppress this entropy only partially, pinning the spin configuration for $B_\parallel>>k_BT/(g\mu_B)$.
However, the fluctuations of the valley degree of freedom are left unpinned due to the small orbital moments of the layer-polarized wave functions near the Brillouin zone corners\cite{Hunt2017, Barrera2022}.  
In the fluctuation regime of a spin polarized, but valley symmetric state, a complete suppression of the entropy by the in-plane Zeeman field is expected.

In Fig. \ref{fig:3}a-b, the entropy is highest in the low density quarter-metal regime but is only partially suppressed by the in plane field, consistent with the pinning of fluctuating spins but not valleys. 
For intermediate densities where the ground state is a spin-polarized half metal\cite{Zhou2022}, instead, the entropy of fluctuating spins is completely suppressed.  
Finally, a residual entropy that is not suppressed by in-plane $B_\parallel$ is observed at higher densities consistent with a fluctuating valley degree of freedom. 
We thus associate both the low- and high- density regimes, where the entropy is not fully suppressed by $B_\parallel$, with fluctuating valley \textit{and} spin degrees of freedom.  

One of the most spectacular manifestations of fluctuating spin moments occurs in $^3$He, where the  anomalous spin entropy of the solid phase favors it at high temperature.  In the resulting ` Pomeranchuk effect',  heating the liquid drives a first order transition to a solid. 
Fig. \ref{fig:3}c shows $\kappa$ as a function of $n_e$ for different temperatures. 
A negative trough in $\kappa$ denotes the first order phase transition between a spin- and valley-polarized phase (PIP$_1$) at higher $|n_e|$ and a symmetric 12-fold phase (Sym$_{12}$) at lower $|n_e|$\cite{Zhou2022} (see inset for schematic illustrations of the Fermi surfaces).
The position of the $\kappa$ minimum ($n_{min}$) changes as a function of temperature as shown in Fig. \ref{fig:3}d: below the $T\approx0.8K$, $n_{min}$ shows a slight movement towards the isospin polarized phase, indicating that the Fermi liquid Sym$_{12}$ phase is entropically favored over the ordered ferromagnet, as is expected if the density of states of the Sym$_{12}$ phase is larger.  
However, above this temperature---which coincides with the Curie temperature extracted in Fig. \ref{fig:1}a---this trend is reversed. In this regime, the PIP$_1$ phase is apparently entropically favored over the Fermi liquid, presumably due to the sudden onset of the isospin fluctuations.  

We may estimate the entropy change across this transition from the Clausius-Clapeyron relation 
$$\Delta S=-\Delta \mu \cdot \frac{\partial \ln(n_{min})}{\partial T}.$$ 
Here, $\Delta \mu=\int \kappa \cdot dn$ is the change in chemical potential across the transition, represented graphically in Figs. \ref{fig:3}e-f for $T$ = 20 mK and $T$ = 1.5K, respectively.  
Fig. \ref{fig:3}g shows $\Delta \mu$ as a function of temperature; it saturates to $\Delta \mu\approx$ 0.3 meV; combined with the measurement of $\partial \ln(n_{min})/\partial T\approx$ -0.24/K, this gives $\Delta S\approx 0.8 k_B$ per charge carrier, in agreement with the measurement of the entropy via the temperature dependent chemical potential alone.

A key question raised by our observation concerns the origin of the large entropy in the liquid state, given that theories developed for tBLG\cite{Song2022,Shi2022,Hu2023,Hu2023a,Chou2023,Lau2023,Datta2023,Zhou2024,Ledwith2024}, which rely critically on the moire potential, are not applicable. 
The anomalous entropy occurs at temperatures that are still significantly lower than the interaction energy and the bare Fermi energy. A natural picture that may explain this observation is that due to the strong interactions, every carrier in the liquid is instantaneously nearly localized by the potential formed by its neighbors, and hence isospin exchange interactions are suppressed. The most crisp theoretical example for such phenomenon is the ``spin-incoherent Luttinger liquid'' in one dimension~\cite{Fiete2007}, where the temperature is above the bandwidth of the spin excitations but below the bandwidth of the charge excitations. 
A similar picture has been used to explain the properties of liquid $^3$He and $^4$He at intermediate temperatures~\cite{Andreev1979,Vollhardt1984}, and may apply to RMG as well.

\section{High temperature transport anomalies}

\begin{figure*}
    \centering
    \includegraphics[width = 7in]{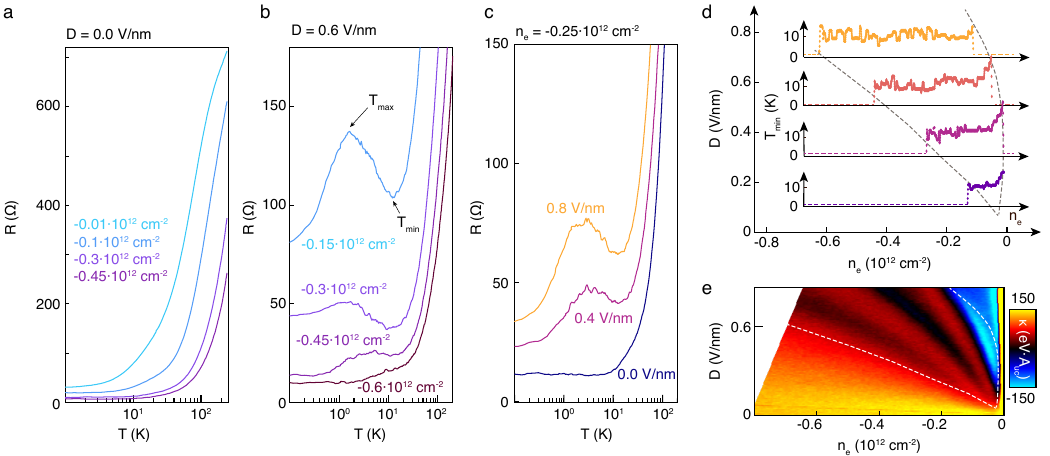}
    \caption{\textbf{Negative temperature coefficients in metallic R2G.} 
    \textbf{(a)} Temperature-dependent resistance in R2G at $D$ = 0.0 V/nm and 
    \textbf{(b)} $D$ = 0.6 V/nm. 
    Several traces show both a finite-temperature resistance maximum (at $T_{max}$ near the Curie temperature) followed by a region where $R(T)$ decreases with temperature, reaching a minimum at  $T_{min}\approx 10K$. 
    \textbf{(c)}  Temperature-dependent resistance at fixed $n_e$= -0.25$ \cdot$ 10$^{12}$ cm$^{-2}$ at different values of $D$.   
    \textbf{(d)} Value of $T_{min}$ as a function of $n_e$ for several values of $D$.  Where no finite temperature minimum is observed, we set $T_{min}=0$.  Finite $T_{min}$ is seen in a  range of the $n_e-D$ plane aligning demarcated by the dashed lines. 
    \textbf{(e)} $\kappa$ measured in the same device at $T=$ 2K.  The dashed lines indicating the region of finite $T_{min}$ align closely with the region of $\kappa$ modulations associated with isospin fluctuations.  
    }
    \label{fig:4}
\end{figure*}

While the thermodynamics of rhombohedral multilayer graphene resemble those of multilayer moir\'e systems, finite temperature transport phenomenology differs dramatically.    
Figure \ref{fig:4}a shows temperature dependent transport in R2G at $D=0$V/nm, where the van Hove singularities are relatively weak and no isospin magnetism is observed. In this regime, $R(T)$ is qualitatively consistent with phonon-dominated scattering\cite{Davis2022}, i.e., approximately temperature independent at low $T$ and increasing for $T\gtrsim 20$ with the onset of phonon scattering. 
Transport behavior is qualitatively distinct at high $D$ (Figs. \ref{fig:4}b-c); here $R(T)$ initial increases up to a $T_{max}$ $\sim 2$K, then \textit{decreases} with increasing $T$ until a finite-temperature resistance minimum is reached at $T_{min} \approx$ 10 to 20 K. 
Notably, the resistance decrease relative to its peak low temperature value can be as large as a factor of three under appropriately chosen experimental conditions (see Figs. \ref{fig:S_ABC_dRdT}, \ref{fig:S_Moire}, and \ref{fig:S_Bpar}).
Above $T_{min}$, the resistance increases rapidly, again consistent with the onset of electron-phonon scattering. Negative $dR/dT$ is observed in the same regions of the  $n_e$- and $D$-tuned parameter space where isospin symmetry-broken phases are observed at low temperatures and increased $\Delta S$ is seen at elevated temperatures (see Figs. \ref{fig:4}d-e). Similar results were obtained for  R3G (Fig. \ref{fig:S_ABC_dRdT}). 
These observations contrast sharply with analogous experiments in tBLG, where the resistivity grows strongly with increasing temperature in the fluctuating moment regime\cite{Polshyn2019,Cao2020,Jaoui2022} and the resistivity is two orders of magnitude higher. 

Negative $dR/dT$ is rare in metals, where most scattering mechanisms are either $T$-independent or increase with temperature.
One possible explanation  is that transport in crystalline layers is described by  electron hydrodynamics at intermediate temperatures\cite{Gurzhi1968}. This scenario is appealing in RMG, since the absence of a moir\'e superlattice and low disorder may lead to slow momentum relaxation.  However, we observe similar negative temperature coefficients in a R2G sample aligned to its hBN substrate (see Fig. \ref{fig:S_Moire}), where the superlattice is expected to introduce strong momentum relaxation via umklapp processes and suppress hydrodynamic effects.  

Alternatively,  negative $dR/dT$ may arise due magnetic fluctuations.  Indeed, resistive anomalies, including regions of negative $dR/dT$ above $T_C$, were observed in elemental ferromagnetic metals such as dysprosium and holmium \cite{Colvin1960}. 
Theoretically, these observations were accounted for\cite{Fisher1968} by invoking scattering of electrons by spin correlations at wave vectors $|q|\approx 2|k_F|$. Within this picture, the decrease in the amplitude of spin correlations with $T$ above $T_C$ accounts for the negative $dR/dT$, although the effects on $R$ are smaller the temperature range of relevance much narrower than observed in our experiment.  
As shown by our entropy measurements, however, short range isospin correlations persist to $T>>T_C$, overlapping with the range of observed negative $dR/dT$. 
Moreover, because $T_C$ is much smaller than the  Bloch-Gruneisen temperature and samples are effectively ballistic at low temperature, negative $dR/dT$ occurs in the absence of a large background from lattice or disorder scattering.  These factors may account for the large magnitude of the resistance drop. 

In this picture, the observed negative $dR/dT$ is the result of the scattering of itinerant electrons by `local moments' formed by spatially confined isospin excitations. However, since the local moment fluctuations are ultimately formed by the same electronic degrees of freedom as the itinerant electrons, the question arises of how the momentum of the electronic fluid is relaxed to give rise to a signature in the electrical resistance. 
Also unanswered is why the high temperature transport in crystalline layers contrasts with twisted graphene multilayers. Possible explanations might include the strong disorder typically present in twisted homo-bilayer systems, the effects of moire superlattice scattering in twisted bilayer graphene, or effects arising from the temperature-dependent effective mass in crystalline multilayers discussed in Ref. \cite{Panigrahi2024}.  
An additional question for future work is the precise nature of the  isospin fluctuations relevant to electron scattering. For example, we find experimentally that $dR/dT$ is not suppressed by the application of large in-plane magnetic fields (Fig. \ref{fig:S_Bpar}), which do however suppress the spin entropy. This suggests that resistivity effects originate in localized valley excitons, rather than pure spin excitations. 
 
In conclusion, our data show that rhombohedral multilayer graphene systems feature a separation of energy scales between the formation of a local isospin order parameter and the onset of long-range order.
From the point of view of the ordered phases---i.e., in the low temperature limit---this hierarchy implies the existence of soft isospin modes in crystalline rhombohedral graphene, analogous to those predicted in tBLG\cite{vafek2020,Khalaf2020,Kumar2021} but unanticipated theoretically in RMG until recent time-dependent Hartree-Fock calculations\cite{Vituri2024,Wolf2024}. 
Viewed from the high temperature side, our data show that the  isospin entropy and Pomeranchuk effects associated with apparent `local moment' behavior can arise in the absence of a lattice on the scale of the Fermi wavelength, which has been fundamental to previous theoretical treatments in moire systems\cite{Song2022,Shi2022,Hu2023,Hu2023a,Datta2023,Chou2023,Lau2023,Zhou2024,Ledwith2024}.
Our results motivate theoretical work to develop a universal framework able to capture the physics of fluctuations in flat band systems with strong Berry curvature in the absence of moir\'e potentials or localized electronic orbitals.

\bibliographystyle{apsrev4-1}
\bibliography{FluctuatingMagnetismbib}

\clearpage
\newpage
\pagebreak

\section{Materials and Methods}

\textbf{Resistance and penetration field capacitance measurements.} We utilize standard lock-in techniques with frequencies from 13 to 47 Hz and AC-currents up to 5 nA to measure the 4-terminal longitudinal resistance of the R2G and R3G samples. 
We employ a low temperature capacitance bridge\cite{Holleis2023} for the penetration field capacitance measurements to extract the inverse compressibility $\kappa$. 
Two AC-excitations are applied to the top gate and reference capacitor with a small amplitude of a few mV to insure that phase transitions are minimally broadened. 
A commercial high electron mobility transistor (HEMT), part number FHX35X, is used as impedance transformer.
At temperatures below 1.5 K, this first stage amplifier is operated at low gain to minimize heating of the sample.
The signal is then amplified by a second stage at 4 K (for details on the capacitance circuit see ref. \cite{Holleis2023}).
At higher temperatures, this is not necessary and the HEMT can be operated at a higher gain without second stage amplification.
From the raw penetration field capacitance data $C$ we extract the inverse compressibility $\kappa$ via  $C = C_TC_B/(C_T+C_B+\kappa^{-1}) \approx \kappa C_T C_B$ where $C_T$, $C_B$ are the geometric capacitance of the sample layer to the top and bottom layer, respectively.

\textbf{Optical sensing of fluctuating isospin order:} 
We employ an optical technique which monitors the energy of the 2s exciton in a WSe$_2$ mono-layer placed in direct contact with an R3G flake\cite{Xie2024} to further explore the hypothesis of fluctuating magnetism above the critical temperature. 
The shift in the exciton energy probes the polarizability of the graphene on length scales comparable to the interlayer distance, making it comparatively insensitive to detailed changes in Fermi surface topology, but highly sensitive to the overall extent of occupied states in momentum space and hence to isospin polarization.  Fig. \ref{fig:S_optical}b shows the reflection contrast near the 2s exciton peak for $D$ = 0.75 V/nm as a function of photon energy and $n_e$.  We observe modulations in the isospin polarization for $n_e<0$, corresponding to hole-like carriers polarized to the R3G layer directly in contact with the WSe$_2$.  From this data, we extract the 2s exciton energy at each carrier density from the local maximum in the slope of the reflection contrast as a function of probe energy. 
The stronger screening with increasing charge carrier density leads to a smooth decreasing background of the 2s exciton energy. To extract the signal due to the isospin transitions, we subtract a constant background from data such as Fig. \ref{fig:S_optical}b and find the relative change in 2s exciton energy with respect to the 2s exciton energy at the charge neutral point of R3G (Fig. \ref{fig:S_optical}d). The result is plotted for a range of temperatures from 2-45 K in Fig. \ref{fig:S_optical}d. 
Evidently, isospin polarization modulations persist to temperature as high as 30 K.
We may thus confidently ascribe the observed high temperature $\kappa$-modulations to $n_e$-tuned changes in the spin- and valley occupation.

\textbf{Extraction of excess entropy:} In the following, we describe the  process used to arrive at the $\Delta S$ presented in the main text (e.g., Fig. \ref{fig:3}a,b) from temperature dependent measurements of the inverse compressibility, $\kappa$ = $\partial \mu/\partial n_e$. 

We measure the capacitance $C$ between a wire connected to the top gate and a wire connected to the bottom gate of the device.  As a result, $C$ also contains a parallel parasitic capacitance contribution, which can be temperature dependent but we assume is not density dependent.  
This change in parasitic capacitance from one temperature to another is comparable to our measured signal, with curves differing by several hundred attofarads (aF) at different temperatures.
Our sensitivity to changes in capacitance at fixed temperature, however, is approximately $20 aF/\sqrt{Hz}$.  The capacitance curves used to generate the data in Figure 3a,b each take approximately one hour to acquire, resulting in a precision with which the overall offset due to systematic offset can be determined of less than 1 aF---far better than measurements of the systematic change in parasitic capacitance between different temperatures.

To accurately extract the entropy from a finite temperature difference, then, we must add or subtract a single, density-independent constant from the measured $\delta \kappa$. 
To determine the value of this offset, we resort to the physics of the sample, specifically the assumption that the electrons are a Fermi liquid at high density.  

Our extraction of the entropy relies on the Maxwell relation
$$\left(\frac{\partial S}{\partial n_e}\right)_{T} = -\left(\frac{\partial \mu}{\partial T}\right)_{n_e}.$$ 
This is justified for small change in $T$, as shown in Fig. \ref{fig:S_smallerTsteps}.  
$\mu(T)$ is determined by integrating $\kappa$ with respect to $n_e$, with measurements at different temperatures used to construct a finite difference approximation to $d\mu/dT$.
We may then extract the entropy from $\delta \kappa$ by (double) integration over $n_e$. We fix the constants of integration be demanding that the entropy be zero on the bounds, one of which is at high density ($n_e\sim -0.6\cdot 10^{12} cm^{-2}$) where quantum oscillations are consistent with a four fold degenerate Fermi surface, and one of is at the lowest density  the where the capacitance measurement is valid (i.e., where the RC time of the sample is still much faster than the measurement frequency), $|n_e|<5\times 10^{10} cm^{-2}$.  

We justify setting the entropy to zero at high density from Fermi liquid theory: for low temperatures, the Fermi surface contribution to the entropy $\Delta S_{FL}$ = $\frac{\pi^2}{3} k_B^2 \rho \Delta T$, where $\rho$ is the density of states, is small - of the order of 10$^{-2} k_B$. A comparison of our experimental data with the theoretically expected Fermi liquid entropy $\Delta S_{FL}$ from single particle band structure calculations is shown in Fig. \ref{fig:S_DeltaS}g for $T$ = 2K and $D$ = 0.5 V/nm.
This value is well below our noise floor and hence, we attribute large changes in $\kappa$ due to the entropy of isospin fluctuations. This assumption is further corroborated empirically by the absence of systematic entropy signal at high density, or low $D$, where no isospin magnetism is observed at low temperatures. This is illustrated in Fig. \ref{fig:S_DeltaS}d,f, where we show the result when the  analysis is performed at higher hole densities outside of the symmetry broken phases; we  find negligible entropy.
This comparison at high densities serves as an estimate of the systematic errors of our entropy measurement of $<<$ 0.1 k$_B$.

\textbf{Extraction of Magnetization at 1.5 K:} Measuring $\Delta M$ is analogous to measuring $\Delta S$, replacing $T$ by the applied magnetic field.  We utilize the Maxwell relation
$$\left(\frac{\partial M}{\partial n_e}\right)_{B} = -\left(\frac{\partial \mu}{\partial B}\right)_{n_e}.$$
First, we take the difference of two high resolution measurements of $\kappa$ at 1.5 K and two different in-plane magnetic fields (Fig. \ref{fig:S_DeltaM} a), b)). 
Integrating the result twice as above, we find the total change in magnetization $\Delta M$ when applying in-plane magnetic fields. 
We compare this result in Fig. \ref{fig:S_DeltaM} d) to the Pauli susceptibility, $\chi_{Pauli} \approx \mu_B^2 \rho(\epsilon_F)$,
where $\mu_B$ is the Bohr magneton, $\rho(\epsilon)$ is the density of states and $\epsilon_F$ the Fermi energy.
The measured susceptibility estimate $\chi$ $\approx$ $\Delta M / $4T = 0.15 $\mu_B$/T exceeds the Pauli susceptibility over a large range of densities.

\textbf{Entropy and the isospin Pomeranchuk effect:} the analysis above on the excess entropy provides the total entropy within each  of the symmetry broken phases.
Here, we determine the entropy difference between a partially polarized and symmetric phase.
R2G undergoes a first order phase transition from a partially isospin polarized phase PIP$_1$  to a symmetric phase Sym$_{12}$ when decreasing $|n_e|$ (fig. \ref{fig:3}c, \ref{fig:S_Pomeranchuk}a).
To determine the entropy difference between PIP$_1$ and Sym$_{12}$, we measure the chemical potential jump at the first order phase transition and the movement of the transition with temperature.
The latter is straight forward and presented in fig. \ref{fig:3} d): $1/n_e \cdot \partial n_e/ \partial T = \partial \ln(n_{min})/\partial T$ = -0.24 $\pm$ 0.02 1/K.
The former is extracted in the following, as illustrated in fig. \ref{fig:S_Pomeranchuk}: first, we perform linear fits to $\kappa$ within PIP$_1$ and Sym$_{12}$ indicated by the green and red dashed lines, respectively, in panels b and c.
We take the deviation from the linear fits beyond 3$\sigma$ as the onset of the phase mixture at the first order phase transition and refer to these density points as $n_{PT}$ in the following.
n$_{PT}$ are marked by the red crosses in panels b and c and are used as integration bounds in the analysis.
Integrating $\kappa$ to get $\Delta \mu$, we need to take three contributions  into account: the change in $\mu$ due to the first order phase transition and the smooth density dependent background of both the low and high density phase due to phase mixing.
As we are only interested in the former, we integrate $\kappa$ and then subtract the mean of the linear contributions of both phases. 
While the transition broadens, the total $\Delta \mu$, decreases with temperature as expected and we find the result presented in the main text Fig. \ref{fig:3} g.

To give an upper bound of the statistical error of $\Delta \mu$ indicated by the red shaded area in Fig. \ref{fig:3} g), we consider two contributions: the error in the slope of the linear fits as well as the uncertainty in determining $n_{PT}$. 
We estimate the latter by adjusting the condition for finding $n_{PT}$ to lie within the grey shaded area in Fig. \ref{fig:S_Pomeranchuk}d) and e). 
The width of the transition changes as a consequence and thus $\Delta \mu$. 
This deviation is taken as the error. 

To determine the chemical potential jump above $T_C$ within error, we take the average $\Delta \mu$ for $T>$1K and propagate the errors accordingly. We find $\Delta \mu$ = 0.3(1) meV.

\textbf{Acknowledgements.} The authors acknowledge  discussions with A. Bernevig, T. Senthil and T. Wang.  
A.F.Y acknowledges the primary support of the National Science Foundation under 
award DMR-2226850. 
C.J. acknowledges support from the Air Force Office of Scientific Research under award FA9550-23-1-0117.
We acknowledge the use of shared facilities 
available due to the support of the National Science Foundation through Enabling Quantum Leap: Convergent Accelerated Discovery Foundries for Quantum Materials Science, Engineering and Information (Q-AMASE-i) award number DMR-1906325. 
The work at MIT was supported by the Science and Technology Center for Integrated Quantum Materials, National Science Foundation grant No.\,DMR-1231319 and was performed in part at Aspen Center for Physics, which is supported by NSF grant PHY-2210452.
K.W. and T.T. acknowledge support from the Elemental Strategy Initiative conducted by the MEXT, Japan (Grant Number JPMXP0112101001) and JSPS KAKENHI (Grant Numbers 19H05790, 20H00354 and 21H05233). 
E.B. was supported by NSF-BSF award DMR-2310312 and by the European Research Council (ERC) under grant HQMAT (grant agreement No. 817799).

\textbf{Author Contributions}
A.F.Y. conceived of and directed the project. 
L.H., T.X., S.X and H.Z. designed and fabricated the devices.  K.W. and T.T. grew the hexagonal boron nitride crystals. 
L.H., C.L.P. and A.F.Y performed the transport and capacitance measurements and analyzed the data. 
B.X., S.X. and C.J. performed the optics measurements.  
L.H., L.S.L., E.B.,  and A.F.Y. wrote the manuscript with input from the other authors.  

\textbf{Data availability}
The experimental data used in this work is available via Zenodo at
https://doi.org/10.5281/zenodo.14641744. The source data is also provided with this article.

\clearpage
\newpage
\pagebreak

\onecolumngrid

\setcounter{equation}{0}
\setcounter{figure}{0}
\setcounter{table}{0}
\setcounter{section}{0}
\makeatletter
\renewcommand{\theequation}{S\arabic{equation}}
\renewcommand{\thefigure}{S\arabic{figure}}
\renewcommand{\thepage}{\arabic{page}}


\newpage
\pagebreak

\begin{figure}[t!]
    \centering
    \includegraphics[width = 120mm]{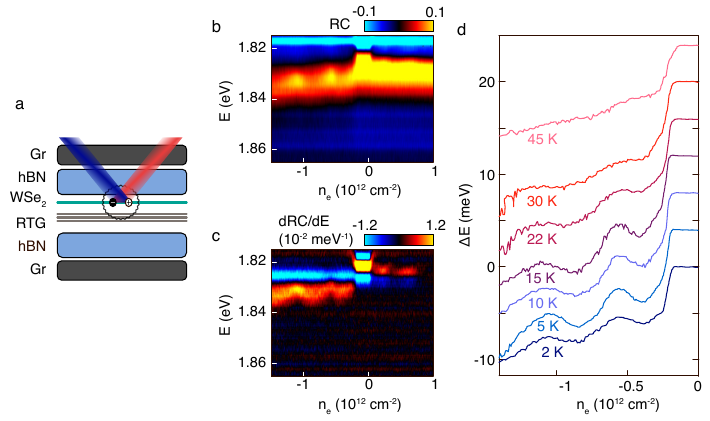}
    \caption{\textbf{Optically detected flavor polarization in R3G.}  
    \textbf{(a)} Measurement schematic of the optical probe technique.  A WSe$_2$ sensing layer is directly in contact with R3G, and its exciton energy is used as a probe of the Fermi surface reconstructions in R3G.  
    \textbf{(b)} Reflection contrast of the R3G/WSe$_2$ heterostructure near WSe$_2$ 2s exiton resonance, as a function of $n_e$ at $D$ = 0.75 V/nm and $T$ = 2 K. 
    \textbf{(c)} Derivative with respect to energy $E$ of the data in panel b.
    \textbf{(d)} Extracted 2s exciton energy at different temperatures. 
    The energy shift $\Delta E$ is measured relative to charge neutrality of the R3G layer, with individual curves offset by 4meV for clarity.
    }
    \label{fig:S_optical}
\end{figure}

\clearpage
\newpage
\pagebreak

\begin{figure*}
    \centering
    \includegraphics[width = 7in]{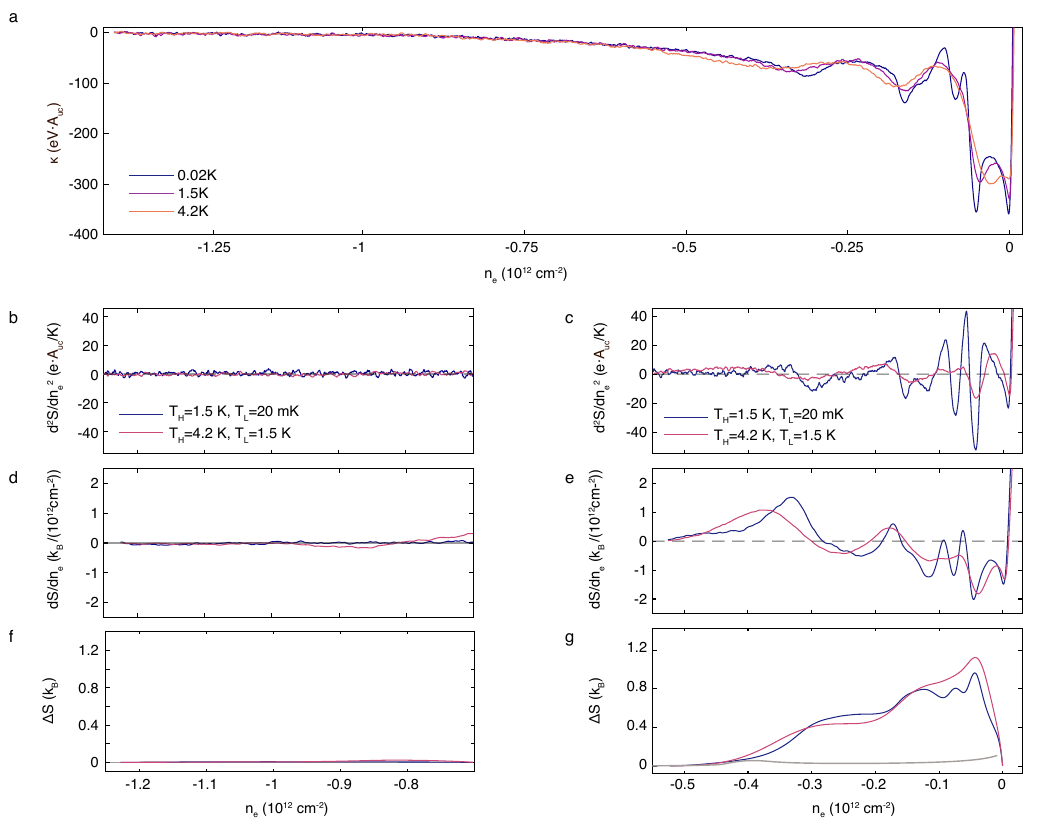}
    \caption{\textbf{Extracting $\Delta S$ from $\kappa(T)$ in R2G.} 
    \textbf{(a)} High resolution measurement of $\kappa$ = $\partial \mu/\partial n_e$ for different temperatures at $D$ = 0.5 V/nm. 
    \textbf{(b)} The difference in $\kappa$ between pairs of temperatures, $\Delta \kappa\cdot(T_H-T_L)^{-1}=(\kappa(T_H)-\kappa(T_L))/\Delta T\approx d^2S/dn_e^2$ measured at high $|n_e|$ where no broken symmetries are observed at lot temperature. 
 \textbf{(c)} $d^2S/dn_e^2$ at in the low-$|n_e|$ regime of broken symmetries. 
 \textbf{(d)} Data from panel b integrated with respect to $n_e$ to give $dS/dn_e$.  
    The bounds of integration are $-1.25\cdot 10^{12}cm^{-2}$ and  $-0.55 \cdot 10^{12}cm^{-2}$. 
\textbf{(e)} Data from panel c    integrated with respect to $n_e$  to give $dS/dn_e$.     The bounds of integration are $
-0.55\cdot 10^{12}cm^{-2}$ and  0, chosen to correspond to the band edge at low $|n_e|$ the onset of appreciable $\Delta \kappa$ at high $|n_e|$. 
\textbf{(f)} Data in panel d integrated with respect to $n_e$ to give $\Delta S$.  The entropy is found to be $\Delta S<<0.1 k_B$ throughout this range. 
\textbf{(g)}  Data in panel e integrated with respect to $n_e$ to give $\Delta S$. This data is shown in the main text as Fig. \ref{fig:3}, panel a. 
The gray lines in both panel f and g show the Fermi liquid entropy $\Delta S_{FL}$ calculated from single particle band structure at finite temperature $T$ = 2K. Evidently, the measured entropy at finite displacement field exceeds the theoretical expectations of a degenerate Fermi liquid by more than an order of magnitude.
Details on this data analysis are given in the methods section.  
    }
    \label{fig:S_DeltaS}
\end{figure*}

\clearpage
\newpage
\pagebreak

\begin{figure*}
    \centering
    \includegraphics[width = 7in]{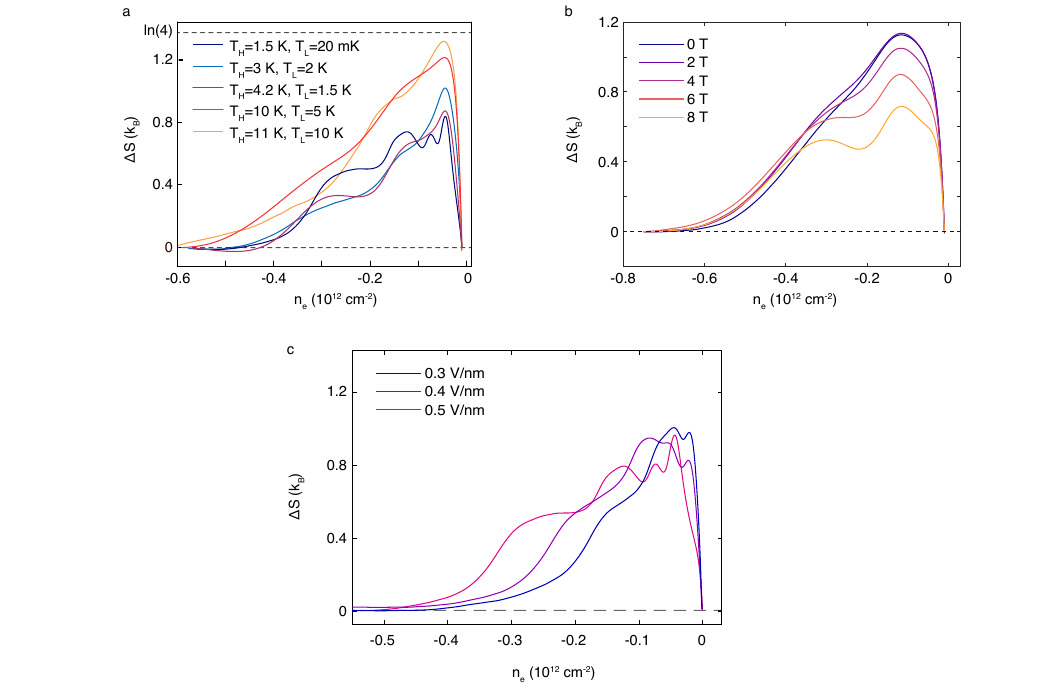}
    \caption{\textbf{Entropy analysis for different temperatures and displacement fields.} 
    \textbf{(a)} Comparison of the measured entropy presented in the main text (Fig. \ref{fig:3}a) with two additional datasets taken at $T_H$ = 3K, $T_L$ = 2K and $T_H$ = 11K, $T_L$ = 10K. Both measurements are in good agreement with the taken over larger temperature ranges, underlyinging that the excess entropy is almost temperature independent.
    \textbf{(b)} In-plane magnetic field dependence of the Entropy at higher temperatures extracted from data taken at $T_H$ = 10K, $T_L$ = 6K. The entropy suppression at higher temperatures is much smaller than in the low temperature data presented in the main text Fig.\ref{fig:3}b, as expected.
    \textbf{(c)} Excess entropy $\Delta S$ for three different displacement fields.
    The entropy is measured for the temperature intervall $T_H$ = 1.5K and $T_L$ = 20mK. The entropy value is found to be approximately independent of the displacement field.}
    \label{fig:S_smallerTsteps}
\end{figure*}

\clearpage
\newpage
\pagebreak

\clearpage
\newpage
\pagebreak

\begin{figure*}
    \centering
    \includegraphics[width = 7in]{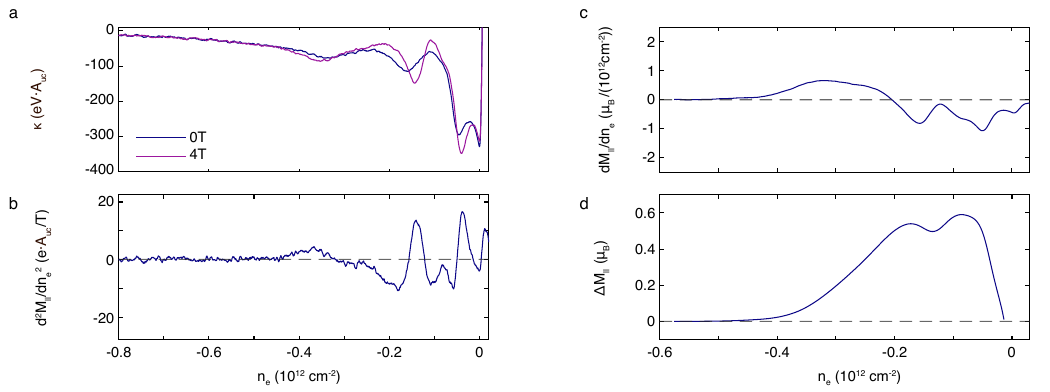}
    \caption{\textbf{Extracting $\Delta M$ from $\kappa(B)$ in R2G.} 
    \textbf{(a)} High resolution measurement of $\kappa$ = $\partial \mu/\partial n_e$ for in-plane fields of 0 and 4T at $D$ = 0.5 V/nm and $T$ = 1.5 K. 
    \textbf{(b)} $ d^2M/dn_e^2=-\Delta \kappa/\Delta B=-(\kappa(B_H)-\kappa(B_L))/\Delta B$ measured for $B_H=4T$ and $B_L=0T$. 
    \textbf{(c)} Data from panel b integrated with respect to $|n_e|$ to give $dM/dn_e$. 
    \textbf{(d)} Data from panel c integrated with resprct to $|n_e|$ to give $\Delta M$.  
    The detailed analysis process is described in the Methods section. 
    }
    \label{fig:S_DeltaM}
\end{figure*}

\clearpage
\newpage
\pagebreak

\begin{figure*}
    \centering
    \includegraphics[width = 7in]{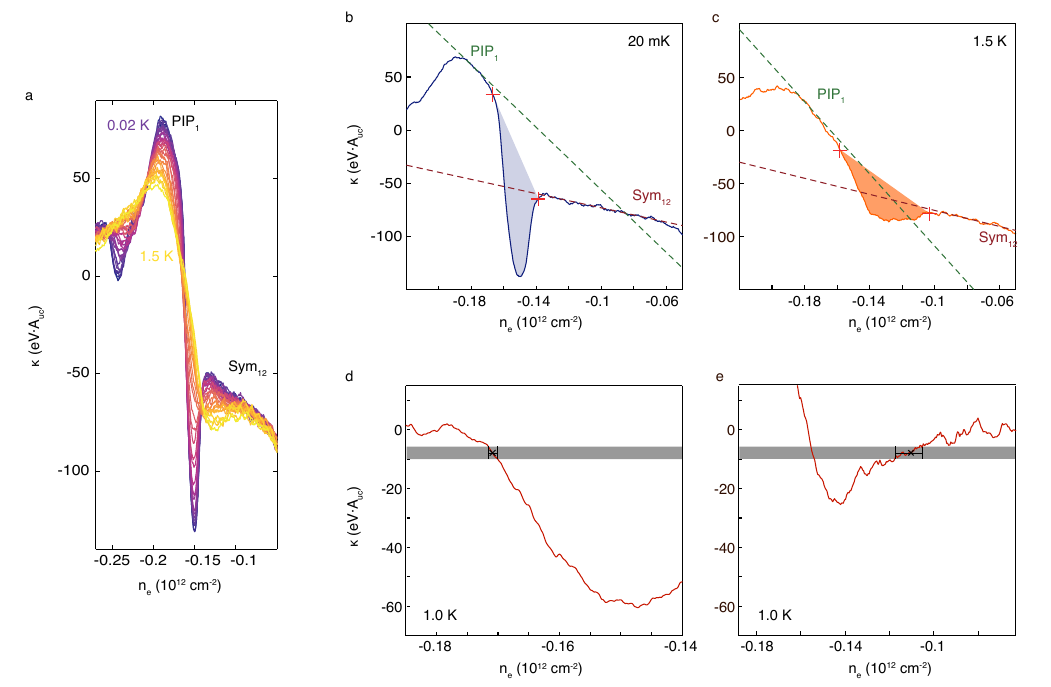}
    \caption{\textbf{Chemical potential discontinuity at the first order phase PIP$_1$ to Sym$_{12}$ transition in R2G.} 
    \textbf{(a)} $\kappa$ at $D$ = 0.8 V/nm in the vicinity of the PIP$_1$ to Sym$_{12}$ transition. 
    \textbf{(b)} $\Delta \mu$ corresponds to the area of the negative dip in $\kappa$. 
    We linearly  extrapolate $\kappa$ in the adjacent phases (dashed lines), and set the bounds of integration where the measured data deviates from the linear trend by more than $3\sigma$, where $\sigma$ is the standard deviation of the measurement. The resulting bounds are indicated by the red crosses. Data here are acquired at 20 mK, with the shaded area corresponding the the measured $\Delta \mu$.  
    \textbf{(c)} The same analysis for data acquired at 1.5 K.  
    \textbf{(d)} $\kappa$ with the trend line of the PIP$_1$ phase subtracted. The grey shaded region is centered at -3$\sigma$, and has a width of $\sigma$; error bars in $\Delta \mu$ are obtained by propagating the uncertainty in determining the bounds which is indicated by the error bar.  
    \textbf{(e)} The same analysis as in panel d), but with the Sym$_{12}$ phase subtracted.}
    \label{fig:S_Pomeranchuk}
\end{figure*}

\clearpage
\newpage
\pagebreak

\begin{figure*}
    \centering
    \includegraphics[width = 7in]{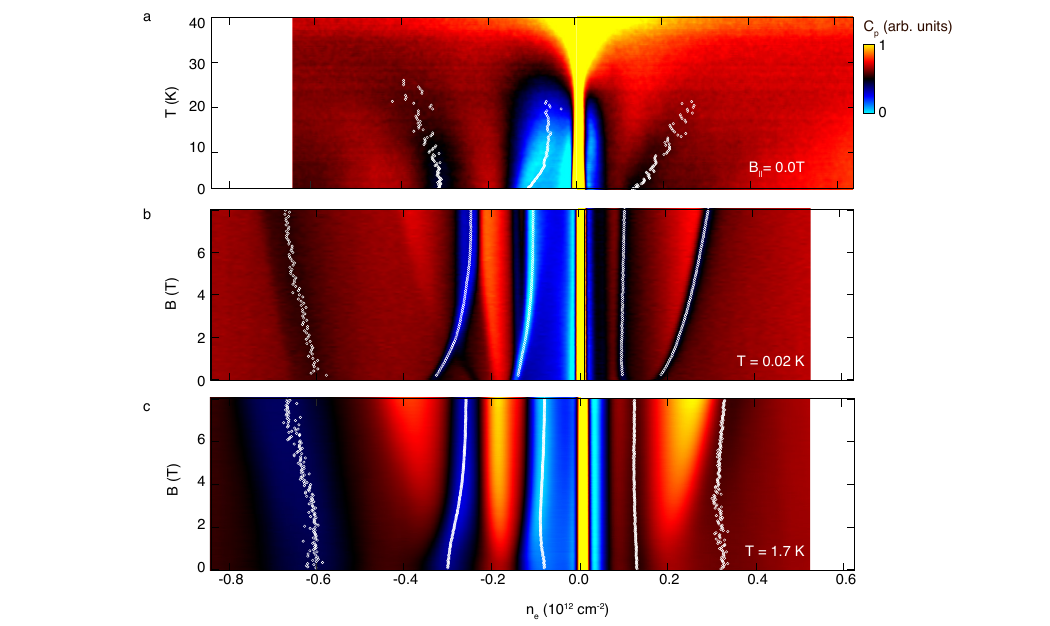}
    \caption{\textbf{Movement of phase transitions with temperature and in-plane field in R2G.} 
    \textbf{(a)} density and temperature dependent penetration field capacitance at $D$ = 0.8 V/nm. 
    \textbf{(b), (c)} in-plane magnetic field dependence at 0.02 K and 1.7 K, respectively, and $D$ = 0.8 V/nm. 
    The white dots mark minima in capacitance which correspond to first order phase transitions at lowest temperatures and show the strong dependence on $T$ and $B_{\parallel}$.
    }
    \label{fig:S_Movement}
\end{figure*}

\clearpage
\newpage
\pagebreak




\begin{figure*}
    \centering
    \includegraphics[width = 7in]{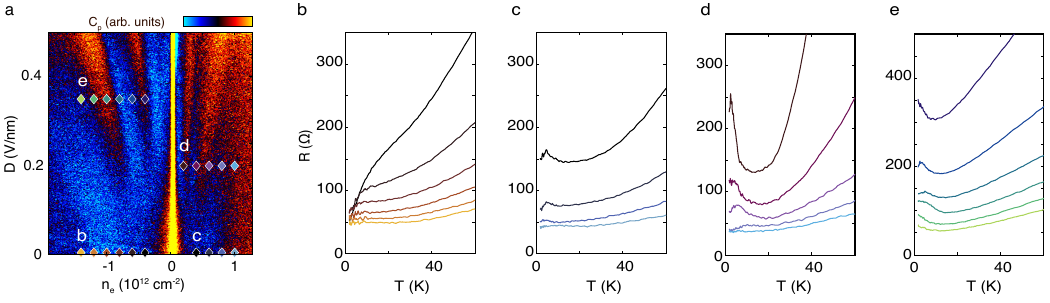}
    \caption{\textbf{Negative $dR/dT$ in R3G.} 
    \textbf{(a)} $n_e$, $D$ phase diagram at $T$ = 6K. 
    \textbf{(b)-(e)} temperature dependent resistance measurements for different densities and displacement fields. The different lines are color coded corresponding to the diamonds in panel a.
    A finite temperature minimum is observed within the fluctuation regime, while such behavior is absent if the ground state is symmetric within the isospin space.
    }
    \label{fig:S_ABC_dRdT}
\end{figure*}

\clearpage
\newpage
\pagebreak

\begin{figure*}
    \centering
    \includegraphics[width = 7in]{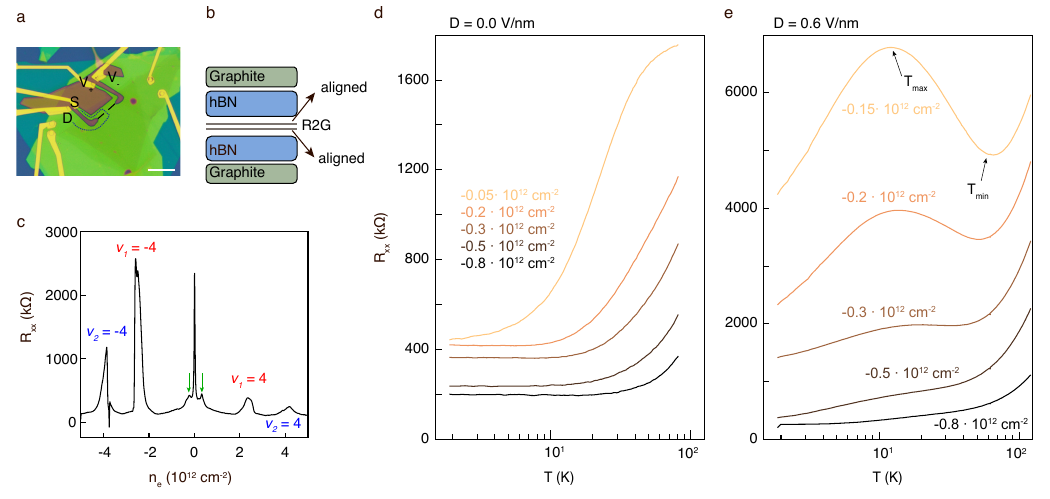}
    \caption{\textbf{Negative $dR/dT$ in sample with hBN-R2G moir\'e sample.} 
    \textbf{(a)} sample micrograph, the scale bar is 5 $\mu m$. The current flow and 4-terminal measurement configuration is indicated. 
    \textbf{(b)} Sample schematic of the double hBN-aligned device.
    \textbf{(c)} $R_{xx}$ as a function of density at $T$ = 1.6 K. The moir\'e peaks of both top and bottom hBN are labelled accordingly, with the green arrows indicating the supermoir\'e peaks at low densities due to double alignment. 
    \textbf{(d)}
    Temperature dependent resistance measurements for decreasing densities at $D$ = 0.0 V/nm and \textbf{(e)} 0.6 V/nm. 
    }
    \label{fig:S_Moire}
\end{figure*}

\clearpage
\newpage
\pagebreak

\begin{figure*}
    \centering
    \includegraphics[width = 7in]{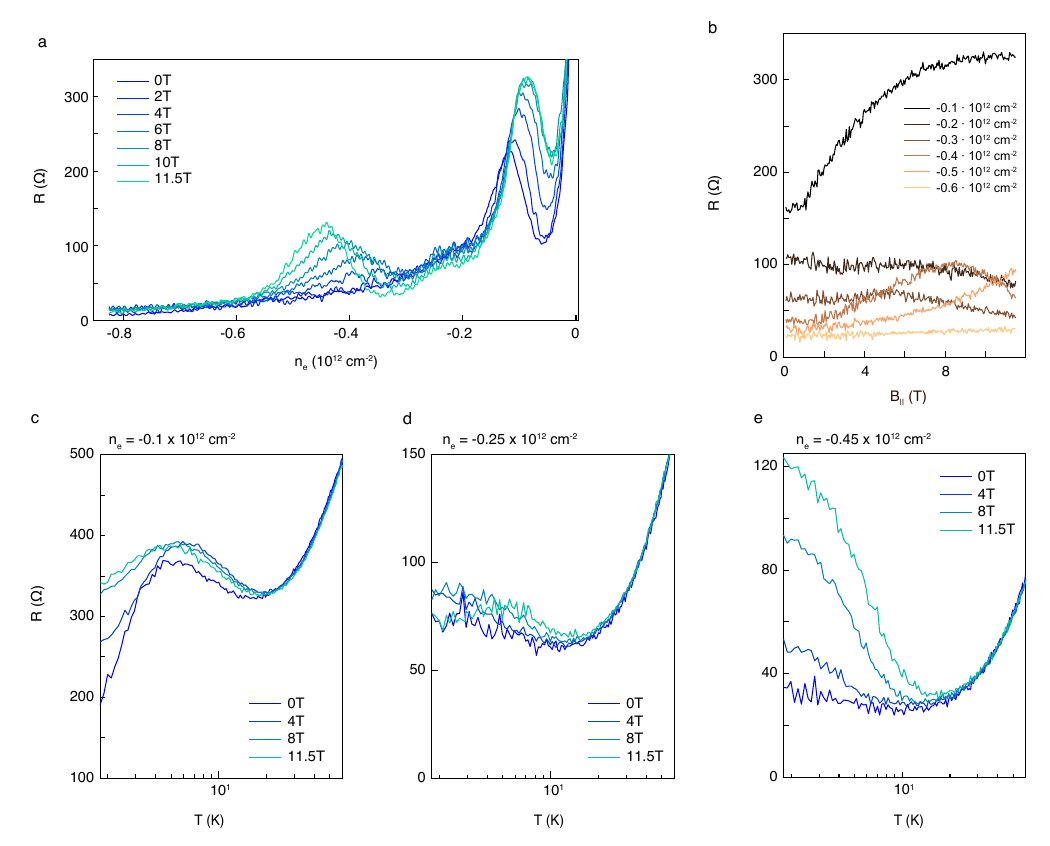}
    \caption{\textbf{Evolution of negative $dR/dT$ with in-plane field in R2G.} 
    \textbf{(a)} Density dependent resistance measurements at 1.7 K and different in-plane magnetic fields.
    \textbf{(b)} $B_{||}$ dependence at 1.7 K for different densities.
    \textbf{(c)} Temperature dependent resistance data for different values of $B_\parallel$ at $n_e=-0.1\cdot 10^{12}cm^{-2}$, 
     \textbf{(d)} $n_e=-0.25\cdot 10^{12}cm^{-2}$, and 
      \textbf{(e)} $n_e=-0.45\cdot 10^{12}cm^{-2}$. 
    While the finite temperature resistance minimum is almost unaffected by the in-plane magnetic field, larger magneto-resistance is observed below $T_{min}$ depending on the low-temperature phases.
    }
    \label{fig:S_Bpar}
\end{figure*}

\begin{table}[ht]
    \centering
    \begin{tabular}{| c | c | c | c |}
    \hline System & Measurement & Finding & Figure \\ 
    \hline \hline R2G & Inverse Compressibility & Isospin transitions above $T_C$& \ref{fig:1}, \ref{fig:3}, \ref{fig:S_DeltaS}, \ref{fig:S_Movement} \\
    \hline R2G & Anomalous Hall & $T_{C,R2G}\approx$ 1K & \ref{fig:1} \\ 
    \hline R2G & Entropy & Large isospin entropy and Pomeranchuk effect & \ref{fig:3}, \ref{fig:S_DeltaS}\\
    \hline R2G & R(T) & Finite temperature resistance minimum & \ref{fig:4}, \ref{fig:S_Bpar}\\
    \hline R2G + Moir\'e & R(T) &Finite temperature resistance minimum &\ref{fig:S_Moire}\\
    \hline R3G & Inverse Compressibility & Isospin transitions above $T_C$
         & \ref{fig:1} \\
    \hline R3G & Anomalous Hall & $T_{C,R3G}\approx$ 3K & \ref{fig:1}\\
    \hline R3G & R(T) & Finite temperature resistance minimum & \ref{fig:S_ABC_dRdT}\\
    \hline R3G+WSe$_2$ & Exciton Sensing & Flavor Transitions above $T_C$ & \ref{fig:S_optical}\\
    \hline
    \end{tabular}
    \caption{\textbf{Summary of the main experimental findings of this article.}}
    \label{tab:my_label}
\end{table}

\end{document}